# Band-gap switching and scaling of nanoperforated graphene

Haiyuan Chen, Xiaobin Niu*

State Key Laboratory of Electronic Thin Films and Integrated Devices, School of Microelectronics and Solid-State Electronics, University of Electronic Science and Technology of China, Chengdu 610054, PR China

* Address correspondence to xbniu@uestc.edu.cn

**ABSTRACT**

In this paper, a framework of {$w_1$, $w_2$, $R$} classification for constructing the graphene nanomesh (GNM) of zigzag-edged hexagonal nanohole is systematically built. The three integer indexes $w_1$, $w_2$, and $R$ indicate the distances between two neighboring sides of nanoholes in two directions and the nanohole size respectively, which leading to a straightforward gap opening criteria, i.e., $w_1 + w_2 - R = 3n + 1$, $n \in \mathbb{Z}$, steered via DFT band structure calculations. The guiding rule indicates that the semimetallic and semiconducting variation is consistent with a peculiar sequence "010" and "100" ("0"/"1" represent gap closure/opening) with a period of 3 for odd and even $w_1$ respectively. The periodic nanoperforation induced gap sizes agreewith a linear fitting with a smaller $\sqrt{N_{rem}}/N_{tot}$ ratio, while deviates from that when $(w_1 + w_2) < R + 1$. Particularly, the {$p$, 1, $p$} and {1, $q$, $q$} structures demonstrate each unique scaling rule pertaining to the nanohole size only when $n$ is set to zero. Furthermore, the coexistence of Dirac and flat bands is observed for {1, $q$, $q$} and {1, 1, $m$} structures, which is sensitive to the atomic patters.

## INTRODUCTION

Monolayer graphene has distinctive two-dimensional (2D) honeycomb-like structure, which is an ideal building block for construction of multifunctional nanostructures.[1-2] It exhibits enormous prospects of being applicable to various areas such as biological imaging[3], photocatalysis,[4] as well as optoelectronic devices.[5] However, these applications are frustrated by the zero gap between the valence and conduction in pristine graphene. Therefore, band gap opening in graphene is crucial and has turned out challenging; moreover, to obtain a tunable unclosed gap pertaining to structural configurations becomes a significant issue for applications of 2D graphene. Extensive approaches have been used to tackle with this problem including dimensionality reduction tailoring,[6-7] surface modification,[8] heteroatom doping,[9] layered stacking,[10] and exerted to external potential.[11]

In addition to those schemes, however recently, the topic about graphene superlattice with nanoholes,[12-14] also dubbed graphene nanomesh (GNM),[15-23] or graphene antidote lattice (GAL),[24-31] attracts substantial research interest since a non-vanishing gap is introduced in certain unique structures. Due to the effective tunability of intrinsic graphene bang gap, these porous graphene structures possess potential applications in spintronics,[12, 14] thermoelectronics,[23, 27] waveguiding devices,[29, 31] and transistors.[15-19, 32] GNM based field-effect transistors perform some improved electronic properties than sliced graphene nanoribbon (GNR) devices,[16] it could deliver 100 times higher drive currents, and demonstrated comparable tunable ON/OFF ratios than similar individual GNR devices. Thus the semimetallic to semiconducting state transition in porous graphene structures plays vital role in pursuing favorable semiconductor devices.

The production of porous graphene structures have been obtained utilizing various techniques, comprising the organic building blocks, etching, and template assisted method.[15-20, 32] The punched graphene films with sub-nanometer holes and narrow neck width are successfully fabricated. Furthermore, the copolymer lithography[16] and imprint lithography[17] lead to sub-10 nanometer neck width, which can meet the requirements for decreasing feature size and opening a band gap in graphene. To explore the usage of GNM in semiconductor devices, a comprehensive understanding of its electronic properties is highly needed. Prior work did by Pedersen et al.[24] theoretically presented the gapped GNM and gave a simple scaling rule shown in equation (1),

$$Eg = k \times \sqrt{N_{rem}/N_{tot}} \qquad (1)$$

where $N_{rem}$ and $N_{tot}$ are the number of removed carbon atoms and original total atoms in a unite cell respectively. However, this rule is constructed for specific GAL and doesn't work well for large $\sqrt{N_{rem}/N_{tot}}$. Besides, not all the GNM structures possess an open band gap, Ouyang et al.[26] showed that half of their built GNMs are semimetals. Oswald et al.[22] pointed that only one third of GNMs remain gapped state.

Although previous theoretical investigations focused on specific patterns, there are still some interesting questions to be probed for GNMs, such as the complete construction of GNMs structures, gap opening/closing switching rule with respect to structural configurations, probability of gapped states in complete structures, and the preferred structure for scaling band gap. In this letter, we present the first-principles computations based on density function theory (DFT) to reveal these answers. The results have valuable implication for designing semiconducting graphene films; likewise, have enlightening significance for band engineering of other 2D group IV materials, such as silicene and germanene.[33-34]

## RESULTS AND DISCUSSION

First, we will shed light on construction of GNMs structures. Previous studies[22, 26] used two important parameters, nanohole size $R$ and neck width (separation of the holes) $w$, to label the different GNMs structures. These two factors are generally contributed to determining the band gap scaling. Here we consider hydrogen passivated GNMs with zigzag-edged hexagonal holes and use three integer numbers, i.e. neck width $w_1$, $w_2$, and radius of nannohole $R$, to describe various GNMs structures as shown in figure 1 (hydrogen atoms are not shown). The neck width $w_1$ along armchair direction, reflecting the odd-even character of GNMs structures, is selected to build the initial GNM pattern with nanohole size $R$. The blue solid line represents the "zigzag-chain-backbone" and the nanohole sliding along the backbone (blue arrow indicated) until the hole edge merge with one of the red lines along zigzag direction (neck width $w_2$) Then we have the GNM patterns with characterized parameter $\{w_1, w_2, R\}$ as illustrated from left to right in fig. 1.

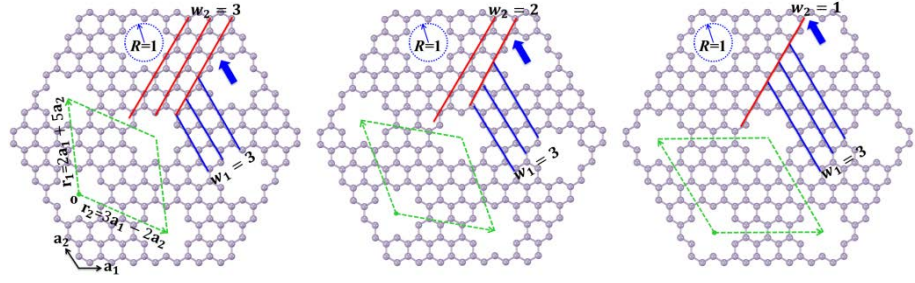

Figure 1. The evolution of $\{w_1, w_2, R\}$ ($w_1$=3) GNMs structures from left to right panel. The blue and red solid lines represent the zigzag chain width of $w_1$ and $w_2$. The holes size $R$ is set to one and the hexagonal unit cell is indicated by green dashed lines.

It is noted that the equivalent structures are obtained if sliding further or along the inverse direction. Thus, one certain GNM structure can be precisely described with $\{w_1, w_2, R\}$ and there is no omissive structure in our built framework. Moreover, the number of configurations $N_{str}$ can be portrayed with equation (2) for different $w_1$ and $R$, and the maximum value of $w_2$ is equal to $N_{str}$ ($1 \leq w_2 \leq N_{str}$).

$$N_{str} = \begin{cases} (w_1 + 2(R-1) + 3)/2, & w_1 \text{ is odd} \\ (w_1 + 2(R-1))/2 + 2, & w_1 \text{ is even} \end{cases} \quad (2)$$

Figure 1 illustrated three GNMs structures, {3, 3, 1}, {3, 2, 1}, and {3, 1, 1}. It is noticed that with defined $w_1$, the number of atoms in unit cell $N_{auc}$ increases inversely when $w_2$ decreases as shown in Table 1. Meanwhile, the $N_{auc}$ for odd and even $w_1$ fulfills the relationship $N_{auc}^n - N_{auc}^{n-1} = 4n - 4$ (n = 2, ..., $w_1/2 + 3/2$), $4n - 6$ (n = 2, ..., $w_1/2 + 2$) respectively.

Table 1. The odd-even character of GNMs ($1 \leq w_1 \leq 8$, $R$=1). The $N_{str}$ and $N_{auc}$ are the number of structures and atoms in the unit cell respectively. The gap states "0" and "1" indicate band gap closure and opening respectively.

| $w_1$ (odd) | 1 | | 3 | | | 5 | | | | 7 | | | | |
|---|---|---|---|---|---|---|---|---|---|---|---|---|---|---|
| $N_{str}$ | 2 | | 3 | | | 4 | | | | 5 | | | | |
| $w_2$ | 2 | 1 | 3 | 2 | 1 | 4 | 3 | 2 | 1 | 5 | 4 | 3 | 2 | 1 |
| $N_{auc}$ | 8, 12 | | 32, 36, 44 | | | 68, 72, 80, 92 | | | | 116, 120, 128, 140, 156 | | | | |
| gap state | 0 | 1 | 0 | 1 | 0 | 0 | 1 | 0 | 0 | 0 | 1 | 0 | 0 | 1 |
| $w_1$ (even) | 2 | | | 4 | | | | 6 | | | | | 8 | | | | | |
| $N_{str}$ | 3 | | | 4 | | | | 5 | | | | | 6 | | | | | |
| $w_2$ | 3 | 2 | 1 | 4 | 3 | 2 | 1 | 5 | 4 | 3 | 2 | 1 | 6 | 5 | 4 | 3 | 2 | 1 |
| $N_{auc}$ | 18, 20, 26 | | | 48, 50, 56, 66 | | | | 90, 92, 98, 108, 122 | | | | | 144, 146, 152, 162, 176, 194 | | | | | |
| gap state | 1 | 0 | 0 | 1 | 0 | 0 | 1 | 1 | 0 | 0 | 1 | 0 | 1 | 0 | 0 | 1 | 0 | 0 |

Utilizing a lattice transform matrix connecting with those three integer numbers, the construction of $\{w_1, w_2, R\}$ system was explained briefly next. The initial two hexagonal lattice vectors $\mathbf{r}_1$ and $\mathbf{r}_2$ of GNM unit cell can be written as $\mathbf{r}_1 = \mathbf{R}\mathbf{r}_2$, where $\mathbf{R}$ is the rotation operator. Considering the lattice as a coordinate system, i.e., $\mathbf{r} = i\mathbf{a}_1 + j\mathbf{a}_2$ with $i, j \in \mathbb{Z}$, the relationship of $\mathbf{r}_1$ and $\mathbf{r}_2$, with a standard choice of lattice vectors $\mathbf{a}_1 = [\sqrt{3}, 0]$ and $\mathbf{a}_2 = [-\sqrt{3}/2, 3/2]$, may be

written as

$$\begin{pmatrix} i_1 \\ j_1 \end{pmatrix} = \begin{pmatrix} \frac{1}{\sqrt{3}}\sin\theta + \cos\theta & -\frac{2}{\sqrt{3}}\sin\theta \\ \frac{2}{\sqrt{3}}\sin\theta & -\frac{1}{\sqrt{3}}\sin\theta + \cos\theta \end{pmatrix} \begin{pmatrix} i_2 \\ j_2 \end{pmatrix}. \qquad (3)$$

(Here the unit length is chosen to be the intrinsic graphene C-C separation). Using an inverse transformation and substituting $\theta$ ($\theta=120°$, between $\mathbf{r_1}$ and $\mathbf{r_2}$) into Eq. (3), a simple mapping between the two integer pairs $(i_1, j_1)$ and $(i_2, j_2)$ is shown in equation (4),

$$\begin{pmatrix} i_2 \\ j_2 \end{pmatrix} = \begin{pmatrix} -1 & 1 \\ -1 & 0 \end{pmatrix} \begin{pmatrix} i_1 \\ j_1 \end{pmatrix}. \qquad (4)$$

Solution of this linear Diophantine equation yields infinite integer pairs. However, aiming to join the $\{w_1, w_2, R\}$ system, the value of $(i_1, j_1)^T$ can be given as follows with inspection,

$$\begin{pmatrix} i_1 \\ j_1 \end{pmatrix} = \begin{pmatrix} w_2 - 1 \\ w_1 + 2R \end{pmatrix}. \qquad (5)$$

Thus, it is easy to know the lattice vectors of unit cell for GNMs with $(i_2, j_2; i_1, j_1)$ transformation. The integer pairs are (3, -2; 2, 5) for {3, 3, 1} structure exemplified in the leftmost panel in fig. 1.

When the geometric scheme of GNMs is established, it is desirable to explore the band gap opening/closure rules. Displayed by figure 2(a), the electronic band structures of $\{w_1, w_2, 1\}$ GNMs are calculated. Here, we choose two samples $w_1=2, 3$, aiming to find the even-odd character of band gap switching rules.

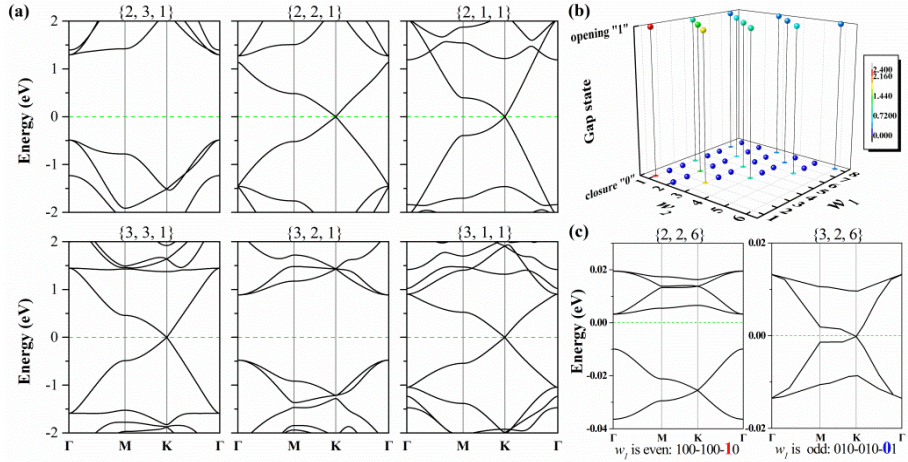

Figure 2. (a) Band structures of $\{w_1, w_2, 1\}$, $w_1=2, 3$; (b) the opened and closed bandgap state of $\{w_1, w_2, 1\}$, indicating by number "0" and "1" respectively, $1<w_1<8$; (c) band structure of {2, 2, 6} and {3, 2, 6}, which consist with the rule.

However, the semimetallic and semiconductor states are both appeared for same $w_1$ whether it is 2 (even) or 3 (odd). Therefore, further computations are needed to discover the rule pertain to the GNM configurations. For such a purpose and avoiding overburdened computational cost, calculations of band structures are steered for $w_1$ from 1 to 8. As the result, the band gap opening/closure states, which is represented by number "1" and "0" respectively, are depicted in fig. 2(b). A simple governing rule, i.e. "010", "100" is found for odd and even $w_1$ respectively. More clearly, the band gap state switching complies with the peculiar cyclic sequence "010 010…" and "100 100…" according to the parity of neck width $w_1$. Taking into account $w_1$ and $w_2$ together, a following band gap opening rule can be briefly derived seen from equation (6), which is similar with those $3p$, $3p+1$, and $3p+2$ ($p$=1, 2, 3…) rule studied in armchair GNRs.[35]

$$w_1 + w_2 - R = 3n + 1, \ n \in \mathbb{Z} \tag{6}$$

The difference is that the occurrence of nonzero energy gap depends on hole size and neck width jointly, which is more complicated than armchair GNRs. Moreover, the gap scalings show different trends when $n$ takes different values, which will be discussed later. When considering the integer pair in eq. (5), an equivalent opening rule is summarized,

$$m_1 + m_2 = 3N, \ n \in \mathbb{Z}, \tag{7}$$

where $N$ is equal to $(n+R)$. Notably, we found that the gap opening rule $Q=3m$ ($m$=1, 2, 3…) for $\{P \times Q\}$ GNM rectangular hole lattice[22] can be interpreted in eq. (6) as well with introducing of $\{w_1, w_2, R\}$ classification. Seen from figure S1 (seen in the Supporting Information), the energy gaps are opened for (3×3) $_{P \times Q}$ and (3×3) $_{P \times Q}$, which is in accordance with {1, 1, 1} and {2, 1, 2} respectively. A more straightforward relationship is that the gap opening exists (gap state "1") if the number of atoms $N_{auc}$ or $N_{tot}$ is divisible by 6 (relevant $N_{auc}$ is shown in table 1). Besides, the $N_{tot}$ could be expressed via $w_1$, $w_2$, and $R$,

$$N_{tot} = 2[(w_2 - 1)^2 + (w_1 + 2R)^2 - (w_2 - 1)(w_1 + 2R)]. \tag{8}$$

To examine the manifestation of the "010" and "100" switching rule, the band structures of two more configurations with larger hole size were calculated depicted in figure 2(c). The nanohole size was set to 6, thus the $N_{str}$ can be easy obtained via eq. (2) when $w_1$ is defined. The gap states of {3, 2, 6} and {2, 2, 6} shown in fig. 2(c) were the seventh out of eight constructions in the sequences for odd and even neck width $w_1$ respectively. The validation of the gap states shows good agreement with the cyclic sequence.

Therefore, the gap opening/closing switching rule pertaining to our geometric scheme is revealed. According to the revealed cyclic sequence "010…" and "100…", the conclusion that gap opening happens in complete structure configurations with a precise probability one third can be drawn. This probability is equal to that studied in GNM systems with rectangular hole arrangement previously[22]. Observably, the half GNMs with closed gaps studied before[26] were already included in our framework, corresponding with the first configurations in the "010…" sequence for odd $w_1$. In our constructed configurations, the energy gap states switching rule and opening probability were explored systematically and thoroughly.

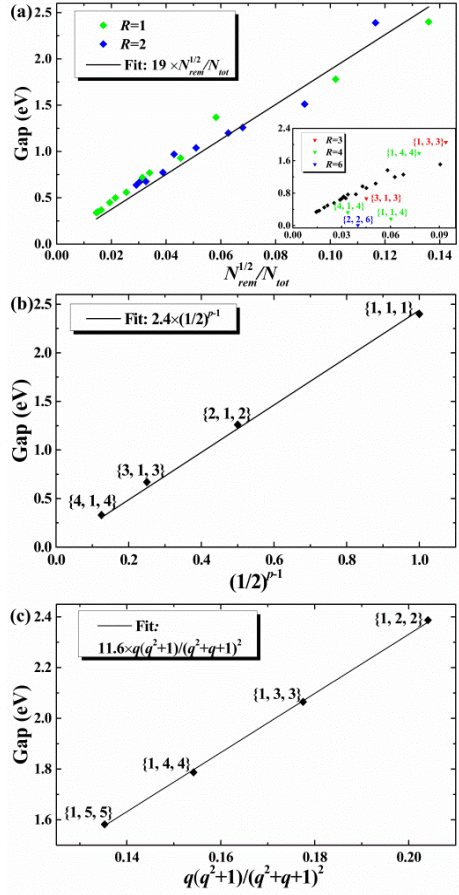

Figure 3. The formulated band gap according to (a) equation (1), (b) equation (9), and (c) equation (10) respectively. Inset of (a) illustrates some larger structures with nanohole radius more than 2.

Next, we will discuss the preferred construction for better gap scaling. Various structures that possibly possessed semiconducting energy gaps were considered to calculate their band structures. Within our framework of $\{w_1, w_2, R\}$ categorization, the $\{w_1, w_2, 1\}$ ($1 \leq w_1 \leq 8$) and $\{w_1, w_2, 2\}$ ($1 \leq w_1 \leq 6$) were calculated. Besides, some larger supercells ($R \geq 2$) were also selected. The selective configurations that satisfy the opening rule described in eq. (6) and eq. (7) presented energy gaps. This could be attributed to the periodic potential introduced by the atomistic "antidote array". Figure 3(a) demonstrated the fitting relationship in keeping with eq. (1)[24] that presented by Pedersen et al.. It is noticed that the energy gap $E_g$ was proportional to such ratio $\sqrt{N_{rem}}/N_{tot}$ when its value is small. However, there were also some structures deviated from such linear properties, such as {2, 2, 6} and {1, 1, 4} seen from inset of figure 3(a). The band structure of {2, 2, 6} and {1, 1, 4} were illustrated in fig. 2(c) and figure 4(c).

Generally, it was found that the energy gap scaling deviated from the linear fitting described via eq. (1) when $(w_1 + w_2) - R < 1$, i.e., n is a negative integer in eq. (6) which was correlated with the hole size and neck width together. On the contrary, it was close to the linear fitting with a small ratio of $\sqrt{N_{rem}}/N_{tot}$, when $(w_1 + w_2) - R > 1$, i.e., n is a positive integer in eq. (6). However, when the sum of neck width $w_1$ and $w_2$ was equal to $R+1$, i.e., $(w_1 + w_2) = R + 1$, there were two kinds of unusual structures {*p*, 1, *p*} and {1, *q*, *q*} performed unique band gap scaling rules differs from eq. (1) respectively. Figure 4(a) and (b) demonstrated the energy band structures of {2, 1, 2} and {1, 2, 2}, beneath which were the schematic of corresponding atomic structures. The unit cells of these two types made us to regard them as superstructures of honeycomb symmetry, which consisted of two "superatoms" stitching mutually without sharing common atoms. Given eq. (8), the unit cell of {*p*, 1, *p*} and {1, *q*, *q*} can be described as $C_{12p^2}H_{6p}$ and $C_{6q+6}H_{6q}$ respectively. Seen from our supporting materials, the band structure of {*p*, 1, *p*} (1≤*p*≤4) and {1, *q*, *q*} (2≤*q*≤5) were depicted in figure S2 and S3 respectively. All those structures have opened energy gaps, which corresponded to the opening rule when *n* is equal to zero in eq. (6). For {*p*, 1, *p*} structures, the gap energy decreased with increasing *p*. Notably, the gap size tuned smaller with a 1/2 factor powered by nanohole size seen from fig. S2. Thus, a well fitted relationship is depicted in fig 3(b) for {*p*, 1, *p*} structures, which is written as

$$E_g = 2.4 \times (1/2)^{p-1}. \quad (9)$$

The band gap scaling of {1, *q*, *q*} was also displayed in fig. 3(c), which was also related to the nanohole size only

$$E_g = 11.6 \times \frac{q(q^2+1)}{(q^2+q+1)^2}. \quad (10)$$

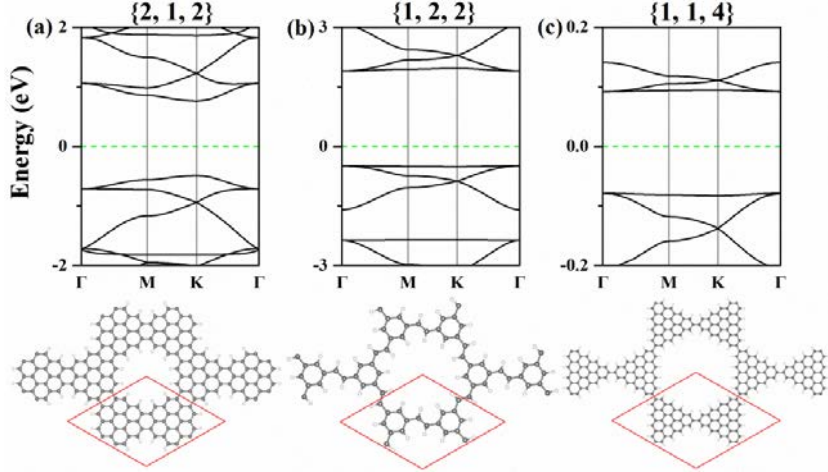

Figure 4. Band structures of (a) {2, 1, 2}, (b) {1, 2, 2}, and (c) {1, 1, 4} respectively, beneath which are the corresponding atomic structures and the red lines indicate the unit cells.

In contrast, it indicates a slower reduction trend of gap size as increasing of *q*. What's more, the coexistence of flat and Dirac bands were observed in {1, *q*, *q*} structures, seen fig 4(b) (*q*=2). With increasing the nanohole size, it is shown that the size of band gap gradually turned smaller and the k-linear dispersion became flatter (seen in fig. S3). Similarly, the occurrence of dispersionless bands was also found in {1, 1, *m*} structures. The band structure of {1, 1, 4} was illustrated in fig. 4(c), others were demonstrated in figure S4 (2≤*m*≤4). The atomic unit cell of {1, 1, *m*} seems like encompassing two triangular quantum dots stitching together. The lattice patterns and interactions between the two superatoms are responsible for the flatness of the nontrivial bands in {1, *q*, *q*} and {1, 1, *m*} structures. However, the gap opening was not always happened for {1, 1, *m*} according to eq. (6), in other words, it also conformed to a period of 3. Nevertheless, the opened gap size of {1, 1, *m*} was very smaller than other structures that comparable with $\sqrt{N_{rem}}/N_{tot}$, which was suggested to avoid when designing semiconducting GNMs. Seen from figure S5, the gap size distribution of some GNMs within the {$w_1$, $w_2$, $R$} space was plotted. Although frustrated by computing capacity, the distribution was not in a relatively large space, the preferred structures may be designed beneath the interface $w_1 + w_2 - R = 1$, which divided the space.

## CONCLUSION

In summary, with introducing the framework of {$w_1$, $w_2$, $R$} categorization, the construction of GNMs with hexagonal unit cell with hydrogen passivated zigzag edge are systematical established in this paper. Meanwhile, a simple band gap opening/closing criteria is revealed, which encompasses the neck width and nanohole size together. Furthermore, the gap opening /closing is in accordance with a periodic sequence, which manifests for odd and even neck width differently. According to the cyclic sequence, there are one third GNMs in our framework may unlock the band gap. Besides, the scaling of gap size agrees with the linear relationship well with a smaller $\sqrt{N_{rem}}/N_{tot}$ ratio. Exceptionally, when $(w_1 + w_2) < R + 1$, a deviation may occur from the linear fitting, which is suggested to be avoided when designing the GNMs. In particular, two special structures of {*p*, 1, *p*} and {1, *q*, *q*} demonstrate each unique scaling rule pertaining to the nanohole size only. Moreover, the coexistence of Dirac and flat bands is observed for {1, *q*, *q*} and {1, 1, *m*} structures, which is sensitive to the atomic patterns. Our results have valuable implication for band engineering of graphene films; likewise, have significance for other 2D group IV materials as well, such as silicene and germanene.

## METHODS

Geometry optimisation and electronic structure calculations were steered within the framework of density functional theory (DFT) implemented in Vienna Ab-initio Simulation Package (VASP)[36]. The generalized gradient approximation (GGA)[37] with the exchange-correlation functional of the Perdew-Burke-Ernzerhof functional (PBE)[38] was employed. The projector-augmented wave (PAW)[39] method was used to describe the ion-electron interaction. The plane-wave cutoff energy is set to be 450 eV, and the convergence criterion in the self-consistency process was set to $10^{-6}$ eV. All structures were relaxed until the residue forces on the atoms became smaller than 0.01 eV/Å. More than 13 Å wide region of vacuum was included along the Z directions to avoid interactions between periodic replicas. A gamma centered meshes of k points, ranging from 3×3 to 7×7 were used to sample the Brillouin zone. All the parameters were carefully tested.

*Acknowledgments*

We acknowledge the financial support from the Recruitment Program of Global Young Experts of China and Sichuan one thousand Talents Plan. The computational resource was supported by TianHe-1(A) at National Supercomputer Center in Tianjin.

# Supporting Information for Band-gap switching and scaling of nanoperforated graphene

**Haiyuan Chen, Xiaobin Niu***

State Key Laboratory of Electronic Thin Films and Integrated Devices, School of Microelectronics and Solid-State Electronics, University of Electronic Science and Technology of China, Chengdu 610054, PR China

* Address correspondence to xbniu@uestc.edu.cn

**S1**. Energy bands of $(3\times3)_{P\times Q}$ and $(3\times3)_{P\times Q}$ configurations, described with {1, 1, 1} and {2, 1, 2} respectively, beneath which are the corresponding atomic structures. The blue (red) lines indicate the width length of zigzag chain.

**S2**. Band structures (top panel) and related atomic structures (bottom panel) for $\{p, 1, p\}$ ($1\leq p\leq 4$) classification, the red lines indicate the large hexagonal unit cell.

**S3**. Band structures (top panel) and related atomic structures (bottom panel) for $\{1, q, q\}$ ($2\leq q\leq 5$) classification, the red lines indicate the large hexagonal unit cell.

**S4**. Band structures (top panel) and related atomic structures (bottom panel) for $\{1, 1, m\}$ ($2\leq m\leq 4$) classification, the red lines indicate the large hexagonal unit cell.

**S5**. Gap size distribution of selected GNMs within $\{w_1, w_2, R\}$ space, divided via the blue surface: $w_1+w_2-R=1$.

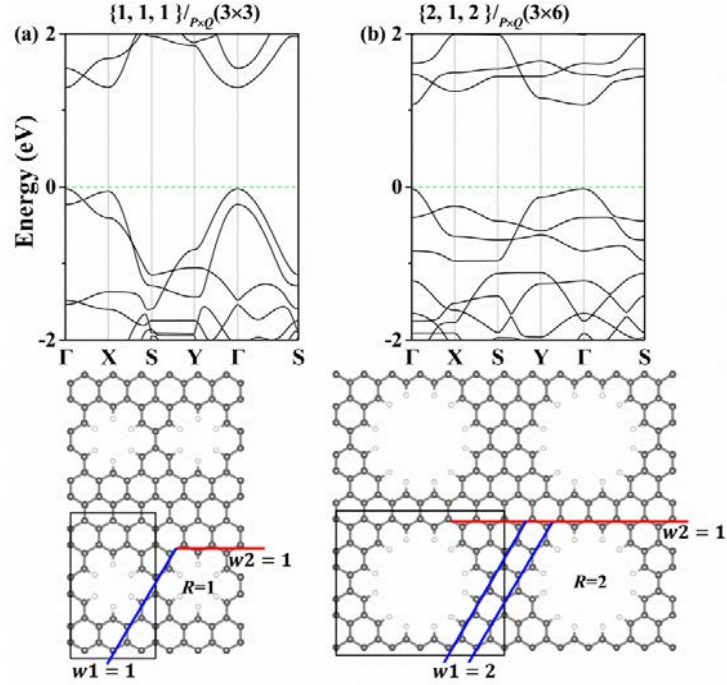

Figure S1. Energy bands of (3×3) $_{P×Q}$ and (3×3) $_{P×Q}$ configurations, described with {1, 1, 1} and {2, 1, 2} respectively.

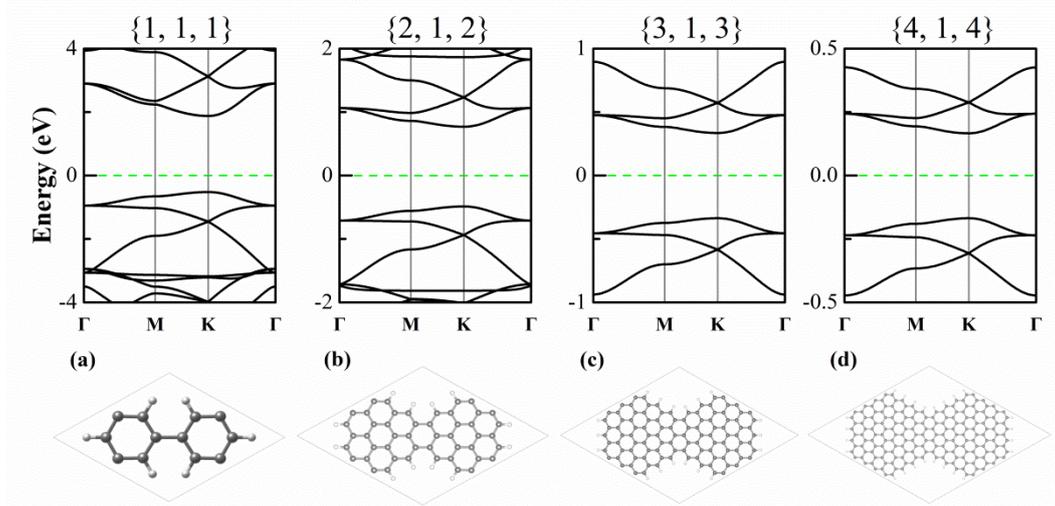

Figure S2. Band structures (top panel) and related atomic structures (bottom panel) for {$p$, 1, $p$} (1≤$p$≤4) classification, the red lines indicate the large hexagonal unit cell.

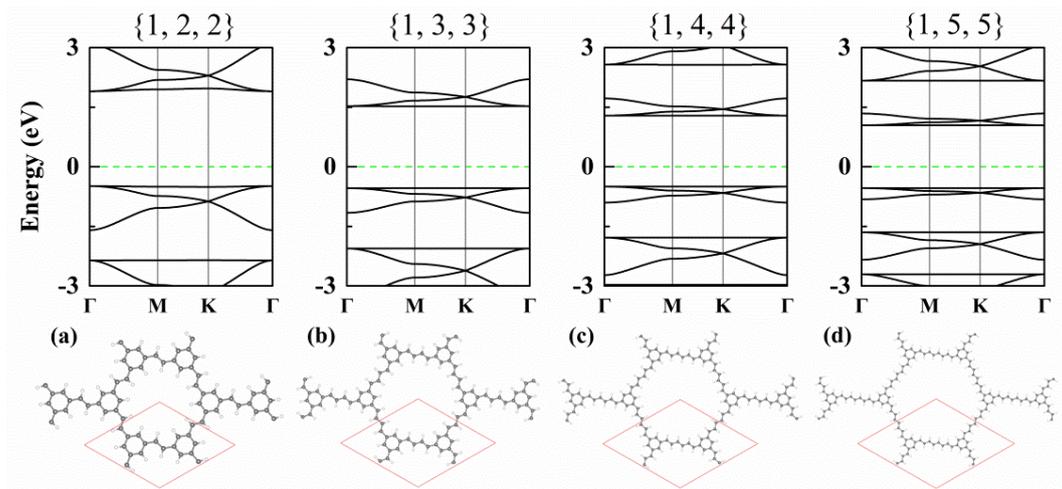

Figure S3. Band structures (top panel) and related atomic structures (bottom panel) for $\{1, q, q\}$ ($2 \leq q \leq 5$) classification, the red lines indicate the large hexagonal unit cell.

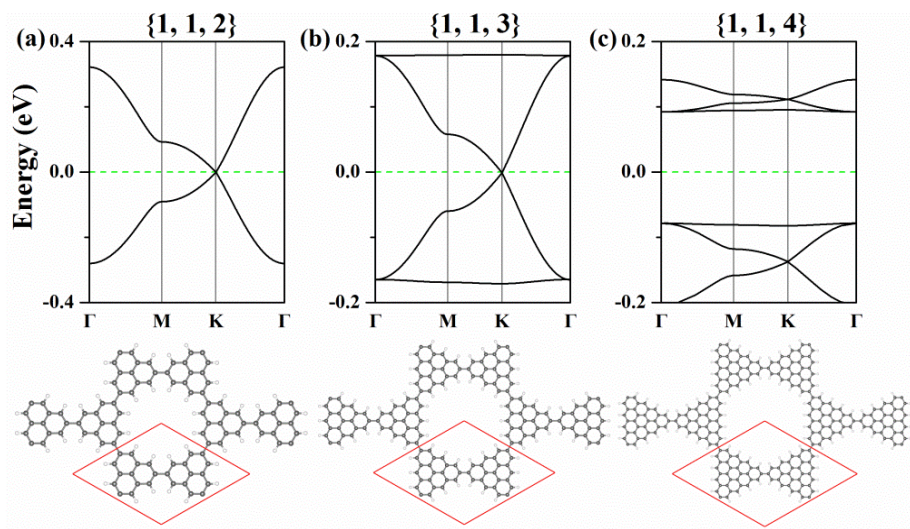

Figure S4. Band structures (top panel) and related atomic structures (bottom panel) for $\{1, 1, m\}$ ($2 \leq m \leq 4$) classification, the red lines indicate the large hexagonal unit cell.

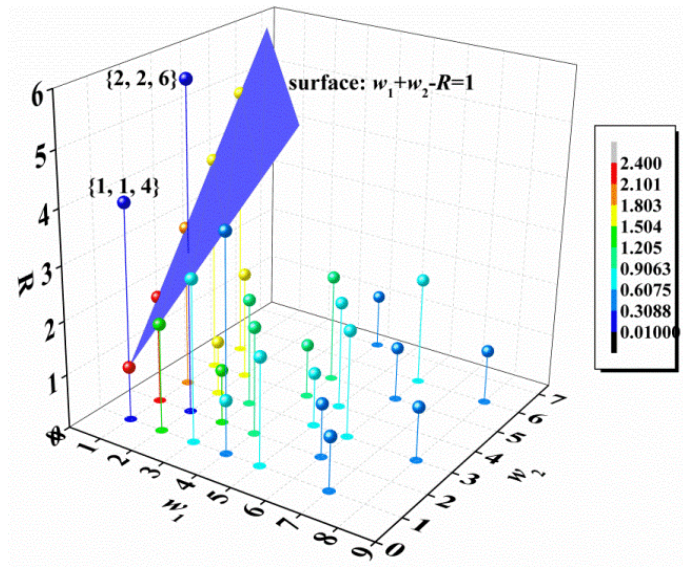

Figure S5. Gap size distribution of selected GNMs within $\{w_1, w_2, R\}$ space, divided via the blue surface: $w_1+w_2-R=1$.